\documentclass{article}

\usepackage[final]{neurips_2019}

\usepackage[utf8]{inputenc} 
\usepackage[T1]{fontenc}    
\usepackage{hyperref}       
\usepackage{url}            
\usepackage{booktabs}       
\usepackage{amsfonts}       
\usepackage{nicefrac}       
\usepackage{microtype}      
\usepackage{amsmath,amssymb}
\usepackage{graphicx}
\usepackage{algorithm}
\usepackage{algpseudocode}
\title{A Multi-Semantic Metapath Model for Large Scale Heterogeneous Network Representation Learning}

\author{Xuandong Zhao\textsuperscript{\rm 1}, Jinbao Xue\textsuperscript{\rm 2}, Jin Yu\textsuperscript{\rm 2}, Xi Li\textsuperscript{\rm 1}, Hongxia Yang\textsuperscript{\rm 2}\\
\textsuperscript{\rm 1}College of Computer Science, Zhejiang University \\     \textsuperscript{\rm 2}Alibaba Group\\
xuandong.zxd@gmail.com, \{zhiji.xjb, kola.yu, yang.yhx\}@alibaba-inc.com, xilizju@zju.edu.cn
} 
\begin{document}

\maketitle

\begin{abstract}
Network Embedding has been widely studied to model and manage data in a variety of real-world applications. However, most existing works focus on networks with single-typed nodes or edges, with limited consideration of unbalanced distributions of nodes and edges. In real-world applications, networks usually consist of billions of various types of nodes and edges with abundant attributes. To tackle these challenges, in this paper we propose a \textbf{m}ulti-\textbf{s}emantic \textbf{m}etapath (\textbf{MSM}) model for large scale heterogeneous representation learning. Specifically, we generate multi-semantic metapath-based random walks to construct the heterogeneous neighborhood to handle the unbalanced distributions and propose a unified framework for the embedding learning. We conduct systematical evaluations for the proposed framework on two challenging datasets: Amazon and Alibaba. The results empirically demonstrate that MSM can achieve relatively significant gains over previous state-of-arts on link prediction.
\end{abstract}

\section{Introduction}
Networks (or graphs) are general data structures to explore and model complex systems of various real-world applications, including social networks, academic networks, physical systems, biological networks and knowledge graphs\cite{hamilton2017representation,kipf2016semi,sanchez2018graph,zhao2019multi}. Mining knowledge in networks has attracted tremendous attention recently due to significant progress in downstream network learning tasks such as node classification, link prediction and community detection.

Network embedding is an effective and efficient way to convert complex network data into a low-dimensional space. Some earlier works have proposed word2vec-based network representation learning frameworks, such as DeepWalk \cite{perozzi2014deepwalk}, LINE \cite{tang2015line}, and node2vec \cite{grover2016node2vec}. They introduce deep learning techniques into network analysis to learn node embedding. However, these works focus on representation learning for homogeneous networks with single-typed nodes and edges. More recently, metapath2vec \cite{dong2017metapath2vec} is proposed for heterogeneous networks, but it is designed for simple "heterogeneous networks" which involve multi-type nodes and single-type edges. PMNE \cite{liu2017principled} and MNE \cite{chang2015heterogeneous} are methods targeted at single-type nodes but multi-type edges. Real-world network-structured applications, such as e-commerce platforms, are much more complicated, containing multi-type nodes, multi-type edges and many attributes. GATNE \cite{cen2019representation} focuses on embedding learning for attributed multiplex heterogeneous networks. Its model can capture both rich attributed information and utilize multiplex topological structures from different node types. However, GATNE can just deal with bipartite graphs which contain two types of nodes. In real-world applications, there are more than two types of nodes and the number of nodes in each type may be unbalanced. 

In this paper, to fill this gap, we propose a novel network embedding framework, MSM, that learns the embeddings on general heterogeneous networks based on multi-semantic metapaths. Concretely, we present the multi-semantic metapath-guided random walks to generate heterogeneous neighborhoods for each node in all different edges. In this way, the model can capture the structure and semantic relations in rich neighbour information and extract necessary nodes and edges for the unbalanced network. Experimental results on real-world datasets demonstrate the improvements of our proposed MSM over other state-of-the-art methods.

To summarize, our work makes the following contributions:

(1) We propose a heterogeneous network embedding method to uncover the semantic and structural information of general heterogeneous networks which have multi-type nodes, multi-type edges and nodes attributes. 

(2) We propose multi-semantic metapath-guided random walks to generate heterogeneous neighborhoods and extract critical nodes and edges to handle unbalanced data.

(3) We demonstrate the effectiveness of our proposed MSM on the Amazon book\&movie and Alibaba datasets. 

\section{Method}
\subsection{Preliminary}
We formalize the problem of heterogeneous network embedding and give some preliminary definitions.

A Heterogeneous Network with multi-type nodes, multi-type edges and nodes attributes is defined as a graph $G = (\mathcal{V}, \mathcal{E}, \mathcal{A})$,
$\mathcal{E}=\bigcup_{r \in \mathcal{R} } \mathcal{E}_{r}$, where $\mathcal{E}_r$ consists of all edges with edge type $r \in \mathcal{R}$, and $\left| \mathcal{R} \right|>1$. Each node $v_i \in \mathcal{V}$ has some types of feature vectors. $\mathcal{A}=\left\{\mathbf{x}_{i} | v_{i} \in \mathcal{V}\right\}$ is the set of node features for all nodes, where $\mathbf{x}_{i}$ is the associated node feature of node $v_i$.

The embedding for heterogeneous network is to give a unified low-dimensional space representation of each node $v$ on every edge type $r$, which aims to learn a mapping function $f_r: \mathcal{V}\rightarrow \mathbb{R}^d$ ($d \ll\left|\mathcal{V}\right|$).

A metapath $\rho(\mathcal{V}, \mathcal{P})$ is defined as a path that is denoted in the form of $\mathcal{V}_{1} \stackrel{\mathcal{P}_{12}}{\longrightarrow} \mathcal{V}_{2} \stackrel{\mathcal{P}_{23}}{\longrightarrow} \cdots \mathcal{V}_{t} \stackrel{\mathcal{P}_{t(t+1)}}{\longrightarrow} \mathcal{V}_{t+1} \cdots \stackrel{\mathcal{P}_{(l-1)l}}{\longrightarrow} \mathcal{V}_{l}$ wherein $\mathcal{P}=\mathcal{P}_{12} \circ \mathcal{P}_{23} \circ \cdots \circ \mathcal{P}_{(l-1)l}$ defines the composite relations between node types $\mathcal{V}_{1}$ and $\mathcal{V}_{l}$.

A multi-semantic metapath $\rho(\mathcal{V}, \mathcal{P}, \mathcal{E})$ is defined as a path that is denoted in the form of $\mathcal{V}_{1} \stackrel{T(\mathcal{P}_{12})}{\longrightarrow} \mathcal{V}_{2} \stackrel{T(\mathcal{P}_{23})}{\longrightarrow} \cdots \mathcal{V}_{t} \stackrel{T(\mathcal{P}_{t(t+1)})}{\longrightarrow} \mathcal{V}_{t+1} \cdots \stackrel{T(\mathcal{P}_{(l-1)l})}{\longrightarrow} \mathcal{V}_{l}$. Different from metapath, in multi-semantic metapath, $T(\cdot)$ represents target edge function and returns $\mathcal{E}_i$ in multiplex heterogeneous network.

\begin{figure*}[htbp]
\centering 
\includegraphics[width=\columnwidth]{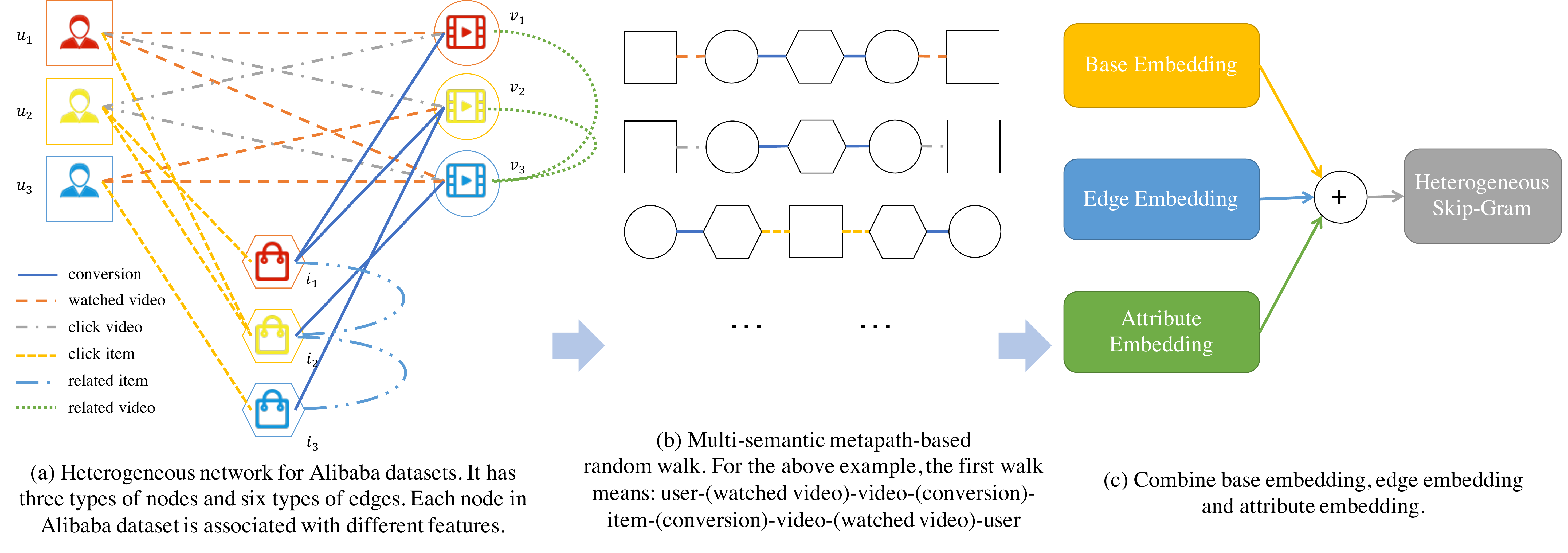} 
\caption{Overview of the multi-semantic metapath model for large scale heterogeneous representation learning --- MSM.}
\label{Fig.pipeline} 
\end{figure*}

For example, Figure\ref{Fig.pipeline}(a) shows the Alibaba dataset with user(U), item(I) and video(V) as nodes. Because there are multiple edges between two different nodes, we can construct multi-semantic metapaths such as (U-watched video-V-related item-I), which can denote that customers like similar items and movies.

\subsection{MSM model}
MSM is designed to cope with networks with heterogeneous and attributed information on both vertices and edges. Following GATNE \cite{cen2019representation}, we propose transductive model and inductive model for MSM. Transductive model can utilize multiplex topological structures from different node types. Inductive model is extended from transductive model, which can capture rich attributed information and generate embeddings on unseen data. The framework of MSM is illustrated in Figure \ref{Fig.pipeline}.

In MSM transductive model (MSM-T), we split the overall embedding of a certain node $v_i$ on each edge type $r$ into two parts: base embedding and edge embedding. The base embedding of node $v_i$ is shared between different edge types. The $k$-th level edge embedding $\mathbf{u}_{i, r}^{(k)} \in \mathbb{R}^{s},(1 \leq k \leq K)$ of node $v_i$ on edge type $r$ is aggregated from neighbors' edge emebeddings as:
\begin{equation}
\mathbf{u}_{i, r}^{(k)}=\sigma\left(\hat{\mathbf{W}}^{(k)} \cdot \operatorname{mean}\left(\left\{\mathbf{u}_{j, r}^{(k-1)}, \forall v_{j} \in \mathcal{N}_{i, r}\right\}\right)\right),
\end{equation}
where $\mathcal{N}_{i, r}$ is the neighbors of node $v_i$ on edge type $r$ and $\sigma$ is an activation function. The initial edge embedding $\mathbf{u}_{i, r}^{(0)}$ for each node $v_i$ and each edge type $r$ is randomly initialized in the model. 

With $K$-th level edge embedding $\mathbf{u}_{i, r}^{(K)}$ denoted as $\mathbf{u}_{i, r}$, we use self-attention mechanism to compute the coefficients $\mathbf{a}_{i, r} \in \mathbb{R}^{m}$ of linear combination of vectors in $\mathbf{U_i}$ on edge type $r$ as:
\begin{equation}
\mathbf{a}_{i, r}=\operatorname{softmax}\left(\mathbf{w}_{r}^{T} \tanh \left(\mathbf{W}_{r} \mathbf{U}_{i}\right)\right)^{T},
\end{equation}
\begin{equation}
\mathbf{U}_{i}=\left(\mathbf{u}_{i, 1}, \mathbf{u}_{i, 2}, \ldots, \mathbf{u}_{i, m}\right).
\end{equation}
where $\mathbf{w}_r$ and $\mathbf{W}_r$ are trainable parameters for edge type $r$ with size $d_a$ and $d_{a} \times s$ respectively. Therefore, the overall embedding of node $v_i$ for edge type $r$ is:
\begin{equation}
\mathbf{v}_{i, r}=\mathbf{b}_{i}+\alpha_{r} \mathbf{M}_{r}^{T} \mathbf{U}_{i} \mathbf{a}_{i, r},
\end{equation}
where $\mathbf{b}_r$ is the base embedding for node $v_i$, $\alpha_r$ is a hyper-parameter denoting the importance of edge embeddings towards the overall embedding and $\mathbf{M}_{r} \in \mathbb{R}^{s \times d}$ is a trainable transformation matrix.

In MSM inductive model (MSM-I), we extend the transductive model to the inductive to handle unobserved data. We update the base embedding and edge embedding: $\mathbf{b}_{i}=\mathbf{h}_{z}\left(\mathbf{x}_{i}\right)$, $\mathbf{u}_{i, r}^{(0)}=\mathbf{g}_{z, r}\left(\mathbf{x}_{i}\right)$. Here $\mathbf{h}_{z}$ and $\mathbf{g}_{z, r}$ are transformation functions such as multi-layer perceptron, where $z$ is node type and $r$ is edge type. 
For the overall embedding of node $v_i$ on type $r$, we also add an extra attribute term:
\begin{equation}
\mathbf{v}_{i, r}=\mathbf{h}_{z}\left(\mathbf{x}_{i}\right)+\alpha_{r} \mathbf{M}_{r}^{T} \mathbf{U}_{i} \mathbf{a}_{i, r}+\beta_{r} \mathbf{O}_{z}^{T} \mathbf{x}_{i},
\end{equation}
where $\beta_{r}$ is a coefficient and $\mathbf{O}_{z}$ is a feature transformation matrix on $v_{i}$'s corresponding node type $z$.

The final transductive and inductive embeddings can be learned by applying multi-semantic metapath-guided random walks. Specifically, given a node $v$ of type $r$ in a MSM walk and the window size $p$, let $v^{-p}, v^{-p+1}, \ldots,$ $v, v^{1}, \ldots, v^{p}$ denote its context. We need to minimize the negative log-likelihood as
\begin{equation}
    -\log P_{\theta_{r}}\left(v^{-p}, \ldots, v^{p} | v\right)=\sum_{1 \leq\left|p^{\prime}\right| \leq p}-\log P_{\theta_{r}}\left(v_{p^{\prime}} | v\right)
\end{equation}
where $\theta_{r}$ denotes all the parameters w.r.t. type $r$ and $P_{\theta_{r}}\left(v_{p^{\prime}} | v\right)$ is defined by the softmax function. The objective function $-\log P_{\theta_{r}}(u | v)$ for each pair of vertices $u$ and $v$ can be easily approximated by the negative sampling method.

In brief, after initializing all the model parameters $\theta$, we generate all training samples \{$(v_i, v_j, r)$\} from multi-semantic metapath-based random walks $\rho_{r}$ on each edge type $r$. Then we train on these samples by calculating $\mathbf{v}_{i,r}$ using Equation (4) or (5) and minimizing negative log-likelihood. We summarize our algorithm in Algorithm 1.

\renewcommand{\algorithmicrequire}{\textbf{Input:}} 
\renewcommand{\algorithmicensure}{\textbf{Output:}}
\begin{algorithm}[htbp]
  \caption{MSM}
  \label{alg:Framwork}
  \begin{algorithmic}[1]
    \Require 
    network $G = (\mathcal{V}, \mathcal{E}, \mathcal{A})$, embedding dimension $d$, edge embedding dimension $s$, learning rate $\eta$, negative samples $L$, coefficient $\alpha, \beta$
    \Ensure 
    overall embeddings $\mathbf{v}_{i,r}$ for all nodes on every edge type $r$
    \State Initialize all the model parameters $\theta$
    \State Generate training samples \{$(v_i, v_j, r)$\} using multi-semantic metapath-based random walks $\rho_{r}$ on each edge type $r$
    \While {not converged}
    \ForAll {$(v_i, v_j, r) \in \text{training samples}$}
    \State Calculate $\mathbf{v}_{i,r}$ using Equation (4) or (5)
    \State Sample $L$ negative samples and update model parameters $\theta$ by minimizing negative log-likelihood Equation (6).
    \EndFor
    \EndWhile
  \end{algorithmic}
\end{algorithm}
 
\section{Experiments}
\subsection{Evaluation Datasets}
For evaluation, we select and adopt two datasets: Amazon and Alibaba. The selection criteria are that they must contain more than two edge types and node types. The details of these datasets are as follows:

The Amazon data used in our experiments is the reviews for movie and book in the Amazon product data\footnote{http://jmcauley.ucsd.edu/data/amazon/}. Customers give 5-score reviews to the product after they finish buying. We divide the data according to that: score 1 and 2 as "dislike", score 3 and 4 as "like" and score 5 as "very like". The statistics of the Amazon dataset is shown in Table \ref{table:alibaba}.

The Alibaba dataset is video-user-item data from Taobao mobile application. There are three types of nodes: user, item and video; and six types of edges. We use \textit{gender}, \textit{city}, \textit{age} and other 3 features as attributes for "user" nodes; \textit{brand}, \textit{price}, \textit{category} and other 6 features as attributes for "item" nodes. For "video" nodes, we extract a 128-dimensional vector from audio through VGG \cite{simonyan2014very} and a 2048-dimensional vector from video through ResNet-50 \cite{he2016deep}. Then, we concatenate these vectors and other 4 features as attributes for "video" nodes. The statistics of nodes and edges in Alibaba datasets is shown in Table \ref{table:alibaba}. We select the interaction data collected during three days as training dataset and use the following day as valid and test dataset. To compare balanced data distribution and unbalanced data distribution, we extract Ali-Balance and Ali-Unbalance datasets. In Ali-Unbalance dataset, there are much more item nodes compared with Ali-Balance dataset. What's more, we vary the remaining edges ratios to test the model's scalability.

\begin{table}[htbp]
\centering
\caption{Statistics of Amazon and Alibaba datasets}
\scalebox{0.9}{
\begin{tabular}{c|c|r|c|rrr}
\hline\hline
      & Types      & Amazon & Types         & \multicolumn{1}{c}{Ali-Balance} & \multicolumn{1}{c}{Ali-Unbalance} & \multicolumn{1}{c}{Ali-Large}  \\ \hline
      & user (U) & 9873 & user (U)            & 10000 & 10000 & 953676        \\
nodes & movie (M) & 5578 & item (I)            & 67068 & 301449 & 4640445        \\
      & book (B) & 4414 & video (V)           & 66306 & 57649  & 1294591       \\
      & Total & 19865 & Total               & 143374 & 369098 & 6888712 \\ \hline
      & dislike (U-M) & 71989 & related item (I-I)  & 173842 & 133925 & 4894802 \\
      & like (U-M) & 74182 & related video (V-V) & 210031 & 92106 & 2459112 \\
      & very like (U-M) & 125189 & click item (U-I)    & 200848 & 279640 & 1723973 \\
edges & dislike (U-B) & 43810 & click video (U-V)   & 192633 & 55188 & 661581 \\
      & like (U-B) & 55564 & watched video (U-V) & 191500 & 36719 & 369448 \\
      & very like (U-B) & 96202 & conversion (I-V)    & 19530 & 79858 & 1275938 \\
      & Total & 466396 &Total               & 988384  & 677436 & 11384854 \\ \hline \hline
\end{tabular}
}
\label{table:alibaba}
\end{table}

\subsection{Baseline Methods}
We compare our model with the following state-of-the-art embedding-based baseline methods and the overall embedding size is set to 200 for all methods.
\begin{itemize}
  \item \textbf{DeepWalk} \cite{perozzi2014deepwalk} applies random walk on the network then uses Skip-gram algorithm to train the embeddings.
  \item \textbf{Node2vec} \cite{grover2016node2vec} adds parameters to control the random walk process and makes it better for certain types of nodes.
  \item \textbf{LINE} \cite{tang2015line} adds link fitting term to the DeepWalk cost function, and samples both one-hop and two-hop neighbors.
  \item \textbf{Metapath2vec} \cite{dong2017metapath2vec} is designed to deal with the node heterogeneity. To compare, we combine all multi-semantic metapaths between the first node and the last node to one path.
  \item \textbf{PMNE} \cite{liu2017principled} proposes three different models to merge multiplex network together to generate one overall embedding for each node. We denote the three methods of PMNE as PMNE(n), PMNE(r) and PMNE(c) respectively.
  \item \textbf{MNE} \cite{chang2015heterogeneous} uses one common embedding and several additional embeddings for each edge type, which are jointly learned by a unified network embedding model. 
  \item \textbf{GATNE} \cite{cen2019representation} proposes both transductive and inductive models for bipartite heterogeneous network embedding.
\end{itemize} 

\begin{table*}[htbp] 
\centering
\caption{Link prediction result for different datasets}
\scalebox{0.8}{
\begin{tabular}{c|ccc|ccc|ccc}
\hline\hline
                & \multicolumn{3}{c|}{Amazon}                      & \multicolumn{3}{c|}{Ali-Balance}                     & \multicolumn{3}{c}{Ali-Unbalance}                    \\ 
Method          & ROC-AUC            & PR-AUC             & F1             & ROC-AUC            & PR-AUC             & F1             & ROC-AUC            & PR-AUC             & F1             \\ \hline
DeepWalk        & 0.780          & 0.782          & 0.710          & 0.786          & 0.729          & 0.720          & 0.711          & 0.701          & 0.655          \\
node2vec        & 0.779          & 0.781          & 0.708          & 0.799          & 0.757          & 0.733          & 0.800          & 0.769          & 0.732          \\
LINE            & 0.697          & 0.678          & 0.640          & 0.728          & 0.705          & 0.674          & 0.708          & 0.671          & 0.655          \\
metapath2vec    & 0.794          & 0.802          & 0.728          & 0.754          & 0.750          & 0.689          & 0.736          & 0.719          & 0.674          \\
PMNE(n)         & 0.783          & 0.782          & 0.712          & 0.786          & 0.811          & 0.747          & 0.758          & 0.734          & 0.700          \\
PMNE(r)         & 0.675          & 0.630          & 0.615          & 0.759          & 0.713          & 0.691          & 0.735          & 0.695          & 0.672          \\
PMNE(c)         & 0.693          & 0.660          & 0.643          & 0.723          & 0.686          & 0.661          & 0.708          & 0.670          & 0.652          \\
MNE             & 0.758          & 0.731          & 0.713          & 0.813          & 0.770          & 0.743          & 0.783          & 0.795          & 0.713          \\
GATNE-T           & 0.825          & 0.794          & 0.746          & 0.837          & 0.831          & 0.770          & 0.814          & 0.815          & 0.741          \\
MSM-T & \textbf{0.871} & \textbf{0.842} & \textbf{0.790} & \textbf{0.866} & \textbf{0.869} & \textbf{0.792} & \textbf{0.857} & \textbf{0.858} & \textbf{0.783} \\ \hline\hline
\end{tabular}
}
\label{table:link}
\end{table*}

\subsection{Experimental Settings}
\subsubsection{Evaluation Metrics}
Because the quality of recommendation services on e-commerce networks can be significantly improved by predicting potential user-to-user or user-to-item relationships, we only test the embedding result on link prediction. Following the commonly used evaluation criteria in similar tasks, we use the area under the ROC curve (ROC-AUC) \cite{hanley1982meaning}, the area under the PR curve (PR-AUC) \cite{davis2006relationship} and F1 score as the evaluation criteria in our experiment. All of these evaluation metrics are uniformly averaged among the selected edge types.

\begin{figure}[htbp]
\centering 
\includegraphics[width=0.6\textwidth]{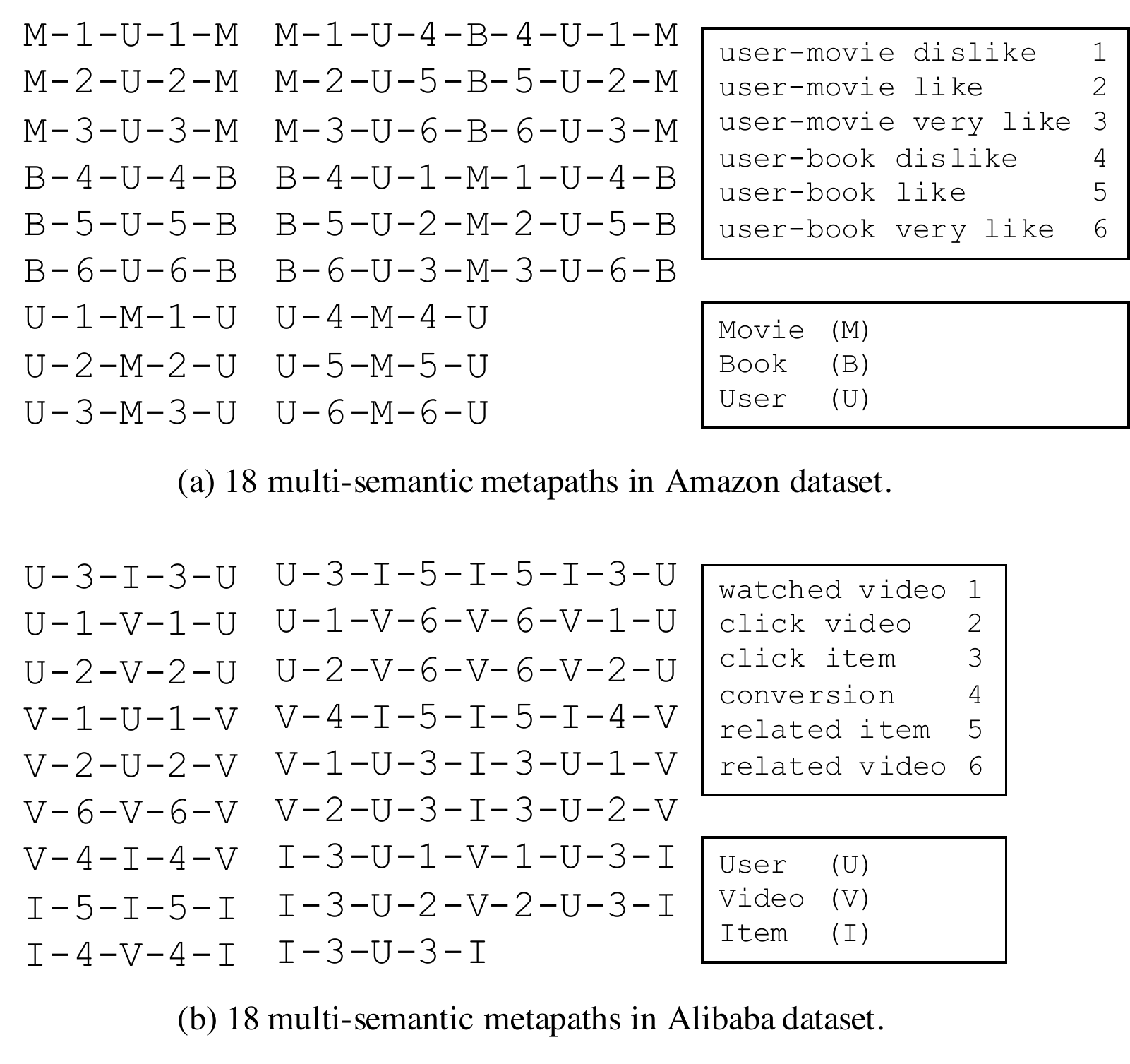} 
\caption{Multi-semantic metapaths in Amazon and Alibaba dataset}
\label{Fig.amazon} 
\end{figure}

\subsubsection{Model Parameters}
To be fair, we set all the embedding dimensions to be 200 in baseline methods and MSM. For all random walk based methods, we set the width of the window to be 10 and select 5 negative samples. For node2vec, we empirically use the best hyper-parameter for training, which is $p = 2$ and $q = 0.5$. For LINE, we set one-hop and two-hop embeddings dimension to be 100 and concatenate them together. For the three PMNE models, we use the hyper-parameters given by their original paper. For MNE, we set the additional embedding size to be 10. We also use the given hyper-parameters in GATNE to train the transductive model (GATNE-T) and inductive model (GATNE-I). For our MSM model, we define several multi-semantic metapaths to do random walks, which is shown in Figure \ref{Fig.amazon}.

\begin{table}[htbp] 
\centering
\caption{Link prediction result for Ali-Large dataset}
\begin{tabular}{c|ccc}
\hline\hline
                & \multicolumn{3}{c}{Ali-Large} \\
Method          & ROC-AUC     & PR-AUC      & F1      \\ \hline
GATNE-I           & 0.719   & 0.691   & 0.722   \\
MSM-I 20\% & 0.726 & 0.774 & 0.733 \\
MSM-I 50\% & 0.732 & 0.755 & 0.737 \\
MSM-I 100\%& 0.746   & 0.748   & 0.782   \\
\hline\hline
\end{tabular}
\label{table:big}
\end{table}

\subsubsection{Quantitative Results} 
\begin{figure}[htbp]
\centering 
\includegraphics[width=0.60\textwidth]{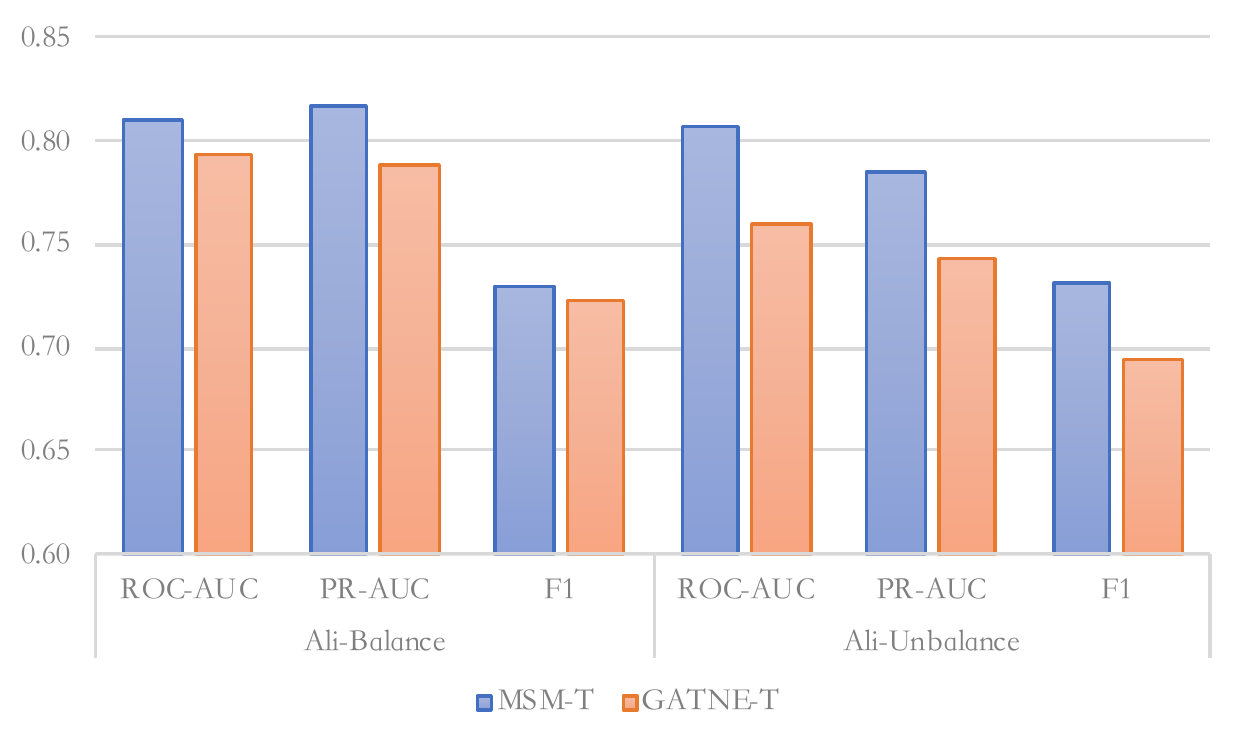} 
\caption{Result for edge "click item" in Ali-Balance and Ali-Unbalance dataset}
\label{Fig.result} 
\end{figure}

The experimental results of three small datasets are shown in Table \ref{table:link}. For each pair of nodes, we calculate the cosine similarity of their embedding. The larger the similarity the more likely there exists a link between them. As the network has more than one relation type, we compute the evaluation metric for each relation type first and take the average of all the relation types as the final result. For the models designed for single layer network, we train a separate embedding for each relation type of the network and use that to predict links on the corresponding relation type, which means that they do not have information from other relation types. We also test one edge ”click item” on both Ali-Balance and Ali-Unbalance datasets to compare MSM with GATNE. The result is shown in Figure \ref{Fig.result}. For Ali-Large dataset, which has much more nodes and edges, the result is shown in Table \ref{table:big}.

The major findings from the results can be summarized as follows: 

(1) MSM outperforms all sorts of baselines in the various datasets. MSM achieves 3.62\% performance lift in ROC-AUC, 8.24\% lift in PR-AUC and 8.31\% in F1-score, compared with best results from GATNE on Ali-Large dataset. 

(2) MSM outperforms GATNE, which demonstrates the importance of adopting multi-semantic metapath-based random walks to construct the heterogeneous neighborhood of a node. 

(3) For Ali-Balance and Ali-Unbalance datasets, MSM performs similarly but baseline methods vary a lot, which shows that MSM is better for unbalanced data distribution.

(4) For Ali-Large dataset, as shown in Table \ref{table:big}, the performances of ROC-AUC and F1 for MSM rise when the percentage of remaining edges increases. When using all edges, MSM outperforms GATNE significantly in large dataset. The results show that MSM can scale for real world heterogeneous networks which contain millions of nodes and edges.

\section{Conclusion}
In this paper, we investigate the representation learning in complex heterogeneous networks which have multi-type nodes, multi-type edges and node attributes. Accordingly, we design a \textbf{m}ulti-\textbf{s}emantic \textbf{m}etapath (\textbf{MSM}) model , which constructs multi-semantic metapath-based random walks to feed into the embedding process. Because MSM can generate balance walks to utilize network structure and node attributes in heterogeneous neighborhood, it is useful in handling unbalanced distribution problems. Extensive experiments and comparisons demonstrate the effectiveness of MSM for representation learning on both fixed networks and unseen nodes. Especially, MSM outperforms representative state-of-the-art embedding approaches on large networks, which shows great value in real-world applications.

\bibliographystyle{unsrt}

\end{document}